\documentclass[]{spie}  

\usepackage[]{graphicx}

\title{A framework for modeling the detailed optical response of thick, multiple segment, large format sensors for precision astronomy applications} 

\author{Andrew Rasmussen\supit{a}, Pierre Antilogus\supit{b}, Pierre Astier\supit{b}, Chuck Claver\supit{c}, Peter Doherty\supit{d}, Gregory Dubois-Felsmann\supit{a}, Kirk Gilmore\supit{a}, Steven Kahn\supit{a}, Ivan Kotov\supit{e}, Robert Lupton\supit{f}, Paul O'Connor\supit{e}, Andrei Nomerotski\supit{e}, Steve Ritz\supit{g} and Christopher Stubbs\supit{d}
\skiplinehalf
\supit{a}SLAC National Accelerator Laboratory, 2575 Sand Hill Rd., Menlo Park, CA 94025, USA; \\
\supit{b}LPNHE/IN2P3/CNRS, UPMC, 4 place Jussieu F75005 Paris, France;\\
\supit{c}National Optical Astronomy Observatory, Tucson, AZ, USA;\\
\supit{d}Department of Physics, Harvard University, 17 Oxford Street, Cambridge, MA 02138, USA;\\
\supit{e}Brookhaven National Laboratory, Upton, NY 11973, USA;\\
\supit{f}Dept. of Astrophysical Sciences, Princeton University, Princeton, NJ 08544, USA\\
\supit{g}SCIPP, University of California at Santa Cruz, Santa Cruz, CA 95064, USA\\
}

\authorinfo{Send correspondence to A.R. -- E-mail: arasmus@slac.stanford.edu; Tel: +1 650 926 2794; Fax: +1 650 926 5566\\
Full electronic version of this paper with color figures is available as arXiv:1407.5655 [astro-ph.IM]}

\begin{document} 
  \maketitle 

\begin{abstract}

Near-future astronomical survey experiments, such as LSST, possess system requirements of unprecedented fidelity\cite{LSST_science_book} that span photometry, astrometry and shape transfer. Some of these requirements flow directly to the array of science imaging sensors at the focal plane. Availability of high quality characterization data acquired in the course of our sensor development program has given us an opportunity to develop and test a framework for simulation and modeling that is based on a limited set of physical and geometric effects. In this paper we describe those models, provide quantitative comparisons between data and modeled response, and extrapolate the response model to predict imaging array response to astronomical exposure. The emergent picture departs from the notion of a fixed, rectilinear grid that maps photo-conversions to the potential well of the channel. In place of that, we have a situation where structures from device fabrication, local silicon bulk resistivity variations and photo-converted carrier patterns still accumulating at the channel, together influence and distort positions within the photosensitive volume that map to pixel boundaries. Strategies for efficient extraction of modeling parameters from routinely acquired characterization data are described. Methods for high fidelity illumination/image distribution parameter retrieval, in the presence of such distortions, are also discussed.
\end{abstract}


\section{INTRODUCTION}
\label{sec:introduction}  

With several large survey imaging instruments either currently in operation (e.g., CFHT/MegaCam, PanStarrs, Suprime-Cam, DECam) or undergoing construction (LSST, others) there is now within the community an intense interest in detailed characterization of the on--sky data, and a generally accepted goal of achieving a deep and quantitative understanding for how CCDs convert incident specific flux into a recorded image of the sky. More specifically, what is needed is a quantitative and extensible expression for the {\it instrument signature}. With such knowledge, underlying models for source flux distribution\footnote{Those source flux distributions are in turn naturally encumbered by system aberrations \& wavefront errors, guiding errors, atmospheric refraction and turbulence, wind buffeting of mirrors, and so on.}  can naturally be constrained. 

In the not-too-distant past, the instrument signature for an imaging instrument may have consisted only of standard ingredients for a CCD data reduction pipeline: a set of dome flats, super flats constructed from data acquired over the night, a set of long-exposure dark frames, a set of representative bias images from which to estimate the instrumental contribution to noise, and so on. The transformation of data from raw to calibrated 
forms\footnote{Absolute calibration would be possible with extractions of standard stars within nearby fields.}
 involved only an arithmetic combination of the preceding data sets, with perhaps some initial suppression of large- and small-scale variations seen in the flat fields.
 In such a reduction process, all small-scale response variations are treated as quantum efficiency (QE) differences and the correction should work perfectly according to this assumption, provided that the spectrum of the flat illumination matches the spectra of the celestial sources to be measured. The sky flat correction provides a correction for illumination that the dome flats cannot provide\footnote{Typical dome flats are approximately Lambertian screens at finite distances and so provide a non-representative illumination pattern at the focal plane.} and this processing step is also appropriate in the limit of zero stray light and ghosting channel contributions. 
  
 The impacts on sky flats by ghosting channels (and any downstream systematic errors) -- clearly deserve a dedicated discussion, but that is outside the scope of this work. From here on we limit our focus to contributions from the sensors themselves.
 
Stubbs\cite{Stubbs_paccd_2014} and others have argued that division by flat field exposures should not be performed as part of a standard image data reduction process step, particularly if any finite part of the photon response non-uniformity (PRNU) can be attributed to non-QE variations. Indeed, as soon as the working assumptions cannot be validated for either of the flat field corrections outlined above, there is a risk that finite systematic errors would be injected, inadvertently, into the calibrated data. This is an undesired byproduct of those processing steps, and the scale of such errors is on the order of the device's PRNU.

There is mounting evidence that current and earlier sensors could be understood better, and in a quantitative way, than they currently are. 
The list that follows is incomplete, and each are discussed further elsewhere\cite{Doherty_spie_2014}, where rich sets of sensor characterization data are analyzed in detail:

\begin{enumerate}

\item {\bf Photon transfer curves (PTC).} Deviations from a linear relationship between {\it variance} and {\it mean} in flat field data sets, those used for generating PTCs. These should nominally exhibit a linear relationship for Poisson statistics. Several years ago, Downing et al.\cite{Downing_2006} proposed that this behavior is a puzzle, and suggested that it be in the interest of the community to better understand it.\footnote{Note that the bend in the photon transfer curve systematically under predicts the gain (ADU/e$^-$) and correspondingly over predicts absolute QEs across the board, whenever system gain is determined using a PTC.}

\item {\bf Brighter-fatter effect.} This shows a correlation\cite{Astier_2013,Antilogus_paccd_2014} between integrated flux and measured widths of the point spread function (PSF) for stellar objects, beginning even at the lowest flux levels.

\item {\bf Anisotropic charge sharing.} Autocorrelations\cite{Antilogus_paccd_2014} of flat field difference images show significant power for single pixel offsets, which turn out to be {\it different by a factor of $\sim$3} for the two pixel address axes (parallel \& serial). While the diagonal term (zero offset) essentially captures the PTC, off-diagonal terms grow faster than the PTC does with flux level.

\item {\bf Tree rings.} Concentric ring-like features in the flat field response correlate with apparent astrometric errors from on-sky star field data\cite{Plazas_paccd_2014,Plazas_pasp_2014}.

\item {\bf Fixed pattern, neighboring pixel response anti-correlations.} Careful analysis\cite{Smith:2008,Kotov:2010} of deep, flat field PRNU maps show different power spectral distribution slopes in the signal sums of contiguous pixels arranged along the two pixel address axes. These studies indicate that a major contributor to the PRNU is traced to small displacements of pixel boundaries - such that a pixel with excess response is very often neighbored by a pixel with deficient response.

\item {\bf Fixed pattern features that outline amplifier segment boundaries and the sensor perimeter.} These are most obvious when full format sensor images are assembled using the data acquired from multiple segments. These include\cite{Doherty_spie_2014} the {\it midline charge redistribution}\footnote{Flat field distortions are seen in the last several rows read out for prototype LSST sensors, next to where the anti-blooming implant is located (between adjacent segments that provide parallel transfer in opposite directions).} and {\it edge rolloff}\footnote{Strong distortions in flat field response are seen around the perimeter of the large format device. In the case of LSST prototype sensors, the distortion is a decrease in flat field signal toward the edge. In the case of DECam, a similar, yet stronger feature is seen with the opposite sign\cite{Plazas_paccd_2014,Plazas_pasp_2014}.}.

\item {\bf Tearing.} Under certain conditions, bimodal, contour-like distortions in the flat field response are seen that apparently migrate in position, from frame to frame, and appear to depend on the illumination history of the sensor. Similar effects have been seen in LBL sensors\cite{Holland_privcom} where the corresponding features were referred to as {\it the city skyline}. While a largely successful search was conducted for sensor operating parameters that would eliminate {\it tearing} (under certain conditions), a physical model for this effect is still desired. There is also a fixed pattern aspect of {\it tearing}, which may appear on the boundaries of adjacent segments on the same side of the anti-blooming implant, even when the {\it tearing} contours are not seen at all in the device interior.

\item {\bf Bamboo.} For some of the prototype sensors for LSST, and for a subset of these that utilized a specific, {\it new} mask fabrication step, a distortion in the flat field response is seen that resembles a bamboo thicket. The pattern is apparently related to the well-known step-and-repeat period of 410$\mu$m.

\item {\bf Spot projector data.} Distributed efforts to generate and analyze some of the effects listed above using {\it spot projector} data\cite{O'Connor_paccd_2014,Tyson_spie_2014} have been crucial in confirming the suspected mechanisms that in turn produce the various flat field distortions. In the analyses performed to date, it was verified that some flat field and astrometric distortions are a consequence of {\it pixel size variations} instead of QE variations.

\item {\bf Backside bias voltage dependence.} A majority of the observed features listed above respond in amplitude to strength of the plane parallel electric drift field near the backside entrance window to the photosensitive bulk. 

\end{enumerate}

In each case, the sensitivity to backside bias voltage dependence is taken as an indication that the underlying mechanism is affected by the ratio of any lateral field strengths to the zero order, plane parallel drift field -- and consequently we know that any distortions in flat field response must be at least partly due to pixel size variations. A corollary to this is that there has to be some depth dependence and so also wavelength dependence, at least where the penetration length into silicon begins to approach relevant scales ($\lambda > 700$nm). We therefore formulate a drift field modeling framework that can be exercised to better understand the relationship between photo-conversion locations within the sensor and the potential wells at the channel where those conversions are collected. Throughout this paper, we use the term {\it flat field distortion} to describe {\it anomalies in the flat field response} where the mechanism for the response anomaly appears
to be due to geometric distortions in the drift field.

\section{SCOPE}
\label{sec:scope}  
\begin{figure}
\begin{center}
\begin{tabular}{cc}
\includegraphics[width=0.3\textwidth]{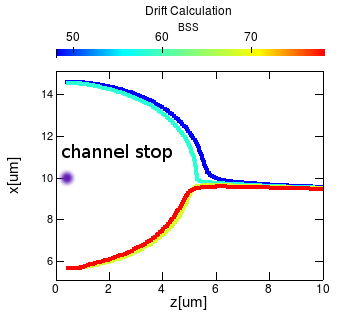}&
\includegraphics[width=0.3\textwidth]{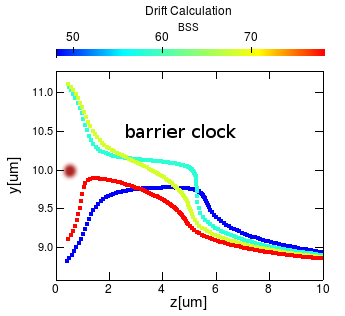}\\
\includegraphics[width=0.3\textwidth]{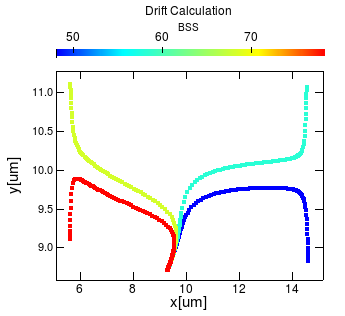}&
\includegraphics[width=0.3\textwidth]{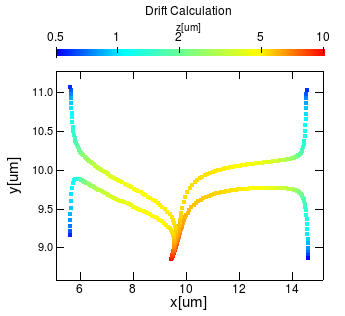}\\
\end{tabular}
\end{center}
\caption
{ 
Example drift calculations with 10$\mu$m periodic, channel stop and barrier clock arrangements with uniform $2D$ dipole moments of $(\vec{\xi}_{\perp CS},\vec{\xi}_{\perp BC})=(-15\xi_0\hat{k},-2\xi_0\hat{k})$. Drift trajectories are distorted by the presence of conversions in the center of pixel $[00]$ - located at $(x,y)=(5,5)\mu\mathrm{m}$ - with an equivalent $3D$ dipole moment of $\vec{p}=-4p_0\hat{k}$. See text for definitions of $\xi_0$ and $p_0$. The starting points for each trajectory on the backside surface were held identical: $\vec{x}_0=(9.31,8.71,100)\mu\mathrm{m}$. Only the one-dimensional {\it backdrop field} is altered between trajectory calculations, by specifying only the backside bias ({\it BSS}). The pixel address of the potential well connecting to $\vec{x}_0$ is then a function of {\it BSS} and four distinct pixels are selected: $[SER,PAR]=([10],[11],[01],[00])$ for $|BSS|=(48,58,68,78)$ volts, respectively. This counterintuitive result is better understood when pixel boundary distortions are considered ({\it cf.} Fig.~\ref{Fig:pixelboundaries}). {\bf Upper left}: the trajectories projected onto the $(x,z)$ plane, where the channel stop axis of symmetry is along the $y$ axis. {\bf Upper right}: the same trajectories projected onto the $(y,z)$ plane. The approximate location of the equivalent barrier is shown in either case, while $\vec{x}_0$ would be located far to the right, at $z=100\mu$m. {\bf Lower left}: the same trajectories projected onto the $(x,y)$ plane. Undistorted pixel boundaries should occur at $x=10\mu$m and $y=10\mu$m and the chosen value for $\vec{x}_0$ is naturally connected to the potential well with address $[SER,PAR]=[00]$. {\bf Lower right}: the same plot as in the lower left, but the color encodes the depth ($z$) of the trajectory, measured from the frontside interface. These figures allow one to determine the depth at which the axial component of the {\it backdrop field} is canceled by the periodic barrier elements. For the parameters used, this apparently occurs at about $z=5\mu$m and $z=1\mu$m for the channel stop barriers and barrier clocks, respectively. At every point along the trajectory, a number of relevant quantities are computed and tabulated in addition to the trajectory itself. These include the electric field strength $|E|$, the electrostatic potential $\phi$, the drift time $t_{coll}$ and the isotropic diffusion $\sigma$. The latter two terms are dependent on the mobility model used\cite{Jacoboni:1977}. Cross-cuts of the periodic potential profile (excluding contributions from collected conversions) are given in Figure~\ref{Fig:potential}.
\label{Fig:driftcalc} 
}
\end{figure}

This work augments an earlier representation of this framework recently published. Rather than duplicate the extensive discussion contained in that paper\cite{Rasmussen_paccd_2014} here, we instead make specific references to sections, equations, figures, etc. -- of that work.  In brief, we developed a computational model that performs a vector line integral, along drift field lines, between points within the photosensitive, depleted silicon bulk and the channel, where the charges are collected. We refer to this as a {\it drift calculation}\footnote{We specifically {\it do not} calculate the complex parallel transfer efficiencies along the parallel address direction, dump transfer into the serial register, or transfer within the serial register toward the readout node. These would be {\it transfer calculations} and are outside the scope of this work.}. The extensive information available by performing the drift calculation is summarized in Figure~\ref{Fig:driftcalc}. Dipole moments are quoted in units of $\xi_0$\footnote{$\xi_0 \equiv 2a\lambda_0 = 10^{-6}\,\mathrm{q}_{e}$ where, according to usual charge dipole moment definition, $2a$ is the linear separation between equal-and-opposite charge concentrations, and $\lambda_0$ is the linear charge density of the charge concentrations.} and $p_0$\footnote{$p_0\equiv2\,z_{ch}\,N\,\mathrm{q}_{e} = 10^5\,\mathrm{q}_{e}\,\mu$m, where again, $z_{ch}$ is the depth of the channel measured from the integrating clock polysilicon equipotential and $N$ is the number of conversions collected in the channel.} for $2D$ and $3D$ terms, respectively. 

In the sections that follow, we briefly describe the foundations of the drift calculation, validate the first order plane parallel field strength, then demonstrate use of the calculation for predicting pixel level distortions, in terms of astrometric, pixel area and (pixel) shape transfer distortions - for a subset of the flat field distortions listed above. We conclude by discussing future directions for matching these calculations to data, and whether pixel level corrections should be performed routinely.

\section{COMPONENTS OF THE MODELING FRAMEWORK} 
\begin{figure}
\begin{center}
\begin{tabular}{cc}
\includegraphics[width=0.3\textwidth]{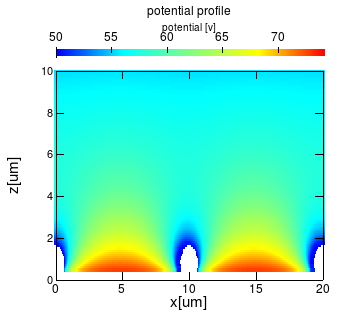}&
\includegraphics[width=0.3\textwidth]{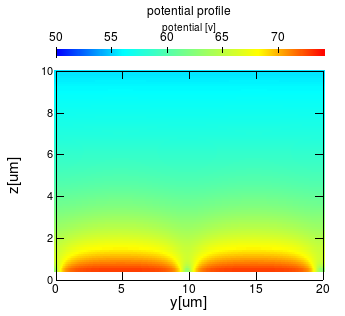}\\
\end{tabular}
\end{center}
\caption
{ 
Electrostatic potential maps in orthogonal planes that also include the center of the pixel with $[SER,PAR]=[00]$.  The vertical axis of both plots is the depth ($z$) measured from the gate structure, and covers roughly 10\% of the sensor thickness. {\bf Left}: the horizontal axis is parallel to the serial address, where the periodic barriers are due to channel stop implants; and {\bf Right}: the horizontal axis is parallel to the parallel address, where barriers are due to barrier clock potentials.  The drift trajectories simply follow the gradient of $\phi$. Figure~\ref{Fig:driftcalc} shows that trajectories with initial positions $\vec{x}_0$ near the backside surface, avoid the volume surrounding the channel stop barriers. The same is true for the barrier clock potentials, where a smaller volume is avoided.
\label{Fig:potential} 
}
\end{figure}

The drift field is represented\cite{Rasmussen_paccd_2014} as a superposition of a 1-D {\it backdrop} field and a periodic arrangement of 
translation-invariant,
linear charge density (channel stop) and dipole (barrier \& integrating clock) elements operating in three dimensions. This choice is natural because at the scale of a pixel volume, clocks and channel stop implants are both linear structures that extend for distances that are many times the scale of an individual pixel, or the thickness of the device. Within the depleted volume, equipotential surfaces due to individual confining potentials are periodic in lateral coordinate and corrugated in the surface normal direction. For the purposes of this study, it should not matter so much that channel stops provide a charge confining barrier field that originates from the bound charge of a depleted p+ implant, while the barrier clocks accomplish a similar, though substantially weaker, field with a distributed arrangement of capacitively coupled surface charges that accumulate at the polysilicon-dielectric interface\setcounter{footnote}{0}\footnote{These free charge surface densities are in turn governed by the shapes of the equipotential volumes of the conductive polysilicon gates in the presence of all other field contributors combined.}. The desired result is that a {\it waffle-like} volume is avoided by drift lines that originate near the backside surface, while the field in the photosensitive bulk is a solution to Poisson's equation -- and this is precisely what is provided by the periodic arrangement of two-dimensional linear charge densities and dipoles as just described. Note that the walls of the waffle that extend in the two perpendicular directions are not of the same height. See Figure~\ref{Fig:potential} for electrostatic potential maps that can be used to visualize these features.

The equipotential boundary conditions, both at the backside entrance window and at the frontside gate structure (if averaged over a full pixel) -- are satisfied by an arrangement of image charges placed outside the bulk of the photosensitive medium, with positions symmetric about the two plane parallel surfaces. The linear charge density of the depleted p+ (channel stop) implant pairs with its opposite-signed image charge to form a two-dimensional dipole in three dimensions, while the barrier clock dipole element also generates an image dipole. The image expansion continues to form a one dimensional, infinite grid of two-dimensional dipoles in three dimensions - that correspond to each of the channel stop or barrier clock confinement fields. The image expansion method is important because it naturally ensures zero in-plane component of the electric field at each of the equipotential planes.

The {\it backdrop field} is in turn a superposition of the {\it built-in field}, governed by the bound charge density within the depleted silicon, and the externally imposed field due to the backside bias. Since the periodic contributions to the field are handled by the dipole image expansion just described, the {\it built-in field} is simply a one-dimensional solution to Poisson's equation\cite{Prigozhin:1998}. We arbitrarily offset the backside bias so that it is identically zero when the electric field strength just inside the backside surface is also zero. Device specific details, such as the operating parameters for an exactly depleted sensor will generally depend on a combination of doping concentration profile and on the actual equivalent dipole field set up by the channel stop implant charge with its image charge expansion - evaluated at the backside surface.

\subsection{The backdrop drift field and its constraints} 
\label{subsec:backdrop}
\subsubsection{$^{55}$Fe X--ray induced charge cloud properties vs. backside bias} 
The backdrop field we use in our drift calculations should be validated. To do this, we use data that can constrain the electric field strength through the photosensitive bulk: $^{55}$Fe X-ray sensor characterization data is available, for a variety of backside bias settings.

As a team, we developed several robust methods\cite{Kotov_spie_2014} to quantify the lateral diffusion suffered by $^{55}$Fe X-ray induced charge clouds during their drift toward the channel. These measurements and analysis methods are also in agreement with sub-pixel optical spot illumination results\cite{Tyson_spie_2014}. We briefly describe the results from one of these methods and compare them to theory based calculations. Starting with a set of pulse height filtered X-ray induced charge cloud islands, we accumulate a specific sum ratio statistic for each X-ray event\footnote{This method is a generalization and expansion of that put forward by Lawrence\cite{Lawrence_pasp_2010}.}. 

We call the sum ratio statistic as ``p4/p9'' -- which refers to the ratio of the central pixel to the sum of the nine pixels including the center. Each event gives a very different statistic, even if the underlying diffusion parameter ($\sigma$) is identical, because of centroid sub pixel coordinate possibilities, partition and readout noises. A given X-ray data set is reduced to a histogram in this sum ratio statistic.

To invert the sum ratio statistic distribution into a maximum likelihood distribution in diffusion ($\sigma$) values, we fit with a response matrix, with the number of fitting parameters equal to the number of rows in the matrix. Rows of the response matrix are generated by computing the probability distribution $f(p4/p9)$ where all possibilities for sub-pixel coordinates, Fano noise, multinomial partition and readout noise are considered - where currently only simple, isotropic Gaussian distributions are considered for the underlying diffused charge distribution. The output of the fitting routine is a ``most probable'' spectrum, $g(\sigma)$. We invoke the assumption that there is a monotonic relationship between the diffusion that the charge cloud suffers and the shallowness of the initial conversion process that liberates the secondary electrons. A standard level is used with the cumulative distribution in $\sigma$, $G(\sigma_{0.96})=0.96$, to represent the maximum diffusion in the data set (which would correspond to surface conversions). 

Diffusion measurement results from multiple segments of the same sensor seem to be in good agreement, and show distinct trends in backside bias voltage: the stronger the electric field, the shorter the collection time and the smaller the diffusion measure. Very good agreement is also achieved between sensors produced by different vendors - which is a desired result, because the underlying medium is high resistivity silicon, which should have universal properties.

\begin{figure}
\begin{center}
\begin{tabular}{c}
\includegraphics[width=0.5\textwidth]{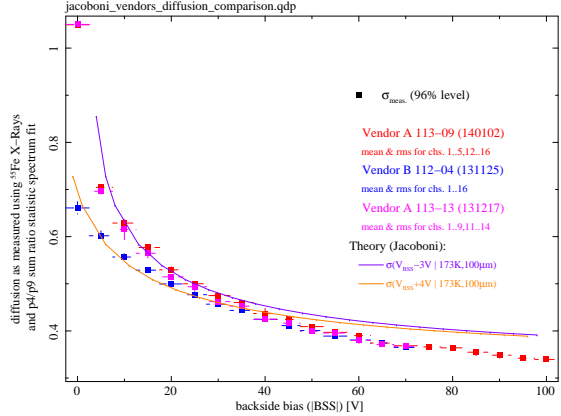}
\end{tabular}
\end{center}
\caption
{ 
Comparison of X-ray charge cloud diffusion measurements to diffusion/drift time calculations of the {\it backdrop field}. The vertical axis is the lateral charge diffusion expressed in pixels (one pixel is equal to 10$\mu$m), and the horizontal axis is the backside bias voltage ($|$BSS$|$). The data taken from two devices from one vendor and one device from another are compared here. Good data was available from most segments of each sensor. The data points and error bars correspond to the mean and rms for the segments available.
\label{Fig:fe55} 
}
\end{figure}

Corresponding diffusion predictions are provided for comparison in Figure~\ref{Fig:fe55}. These include ``eyeballed'' offsets in the bias scale to roughly match up with the two sets of curves corresponding to each vendor. We notice that the measured diffusion continues to decrease as the field strength is increased toward the right, by a significant margin over what the velocity saturation mobility model of Jacoboni\cite{Jacoboni:1977} predicts. Taking this at face value, this would suggest that the Jacoboni model\footnote{Along with other published mobility models.} under predicts the diagonal terms in the mobility tensor for electrons, or that some other mechanism prevents the ($\sim$couple thousand) electron cloud from diffusing in a way that isolated electrons do. We clearly need to re-examine these results with more scrutiny.

\subsubsection{Apparent tree ring flat field distortion amplitude vs. backside bias} 

\begin{figure} 
\centering
\begin{tabular}{cc}
\includegraphics[width=.35\textwidth,angle=0]{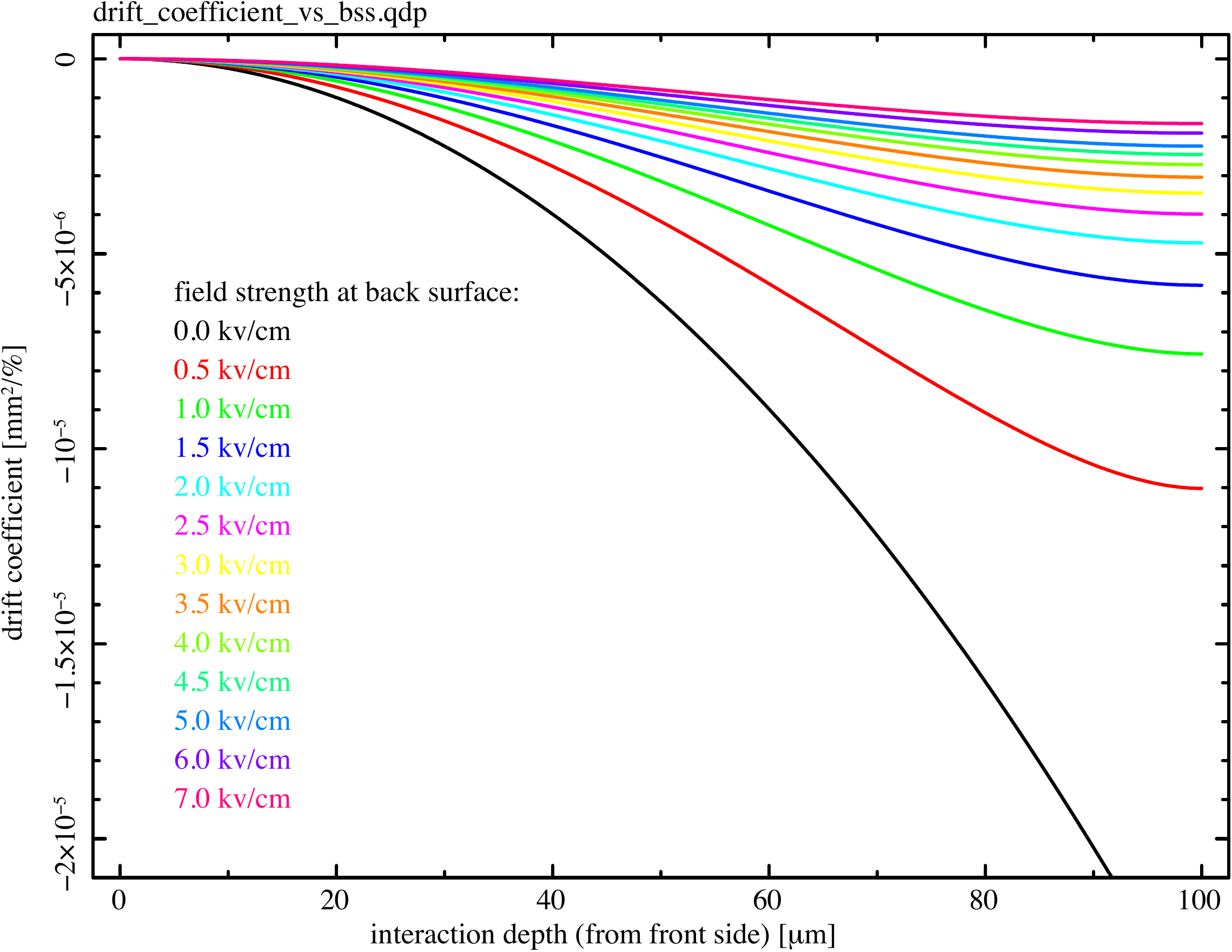}&
\includegraphics[width=.35\textwidth,angle=0]{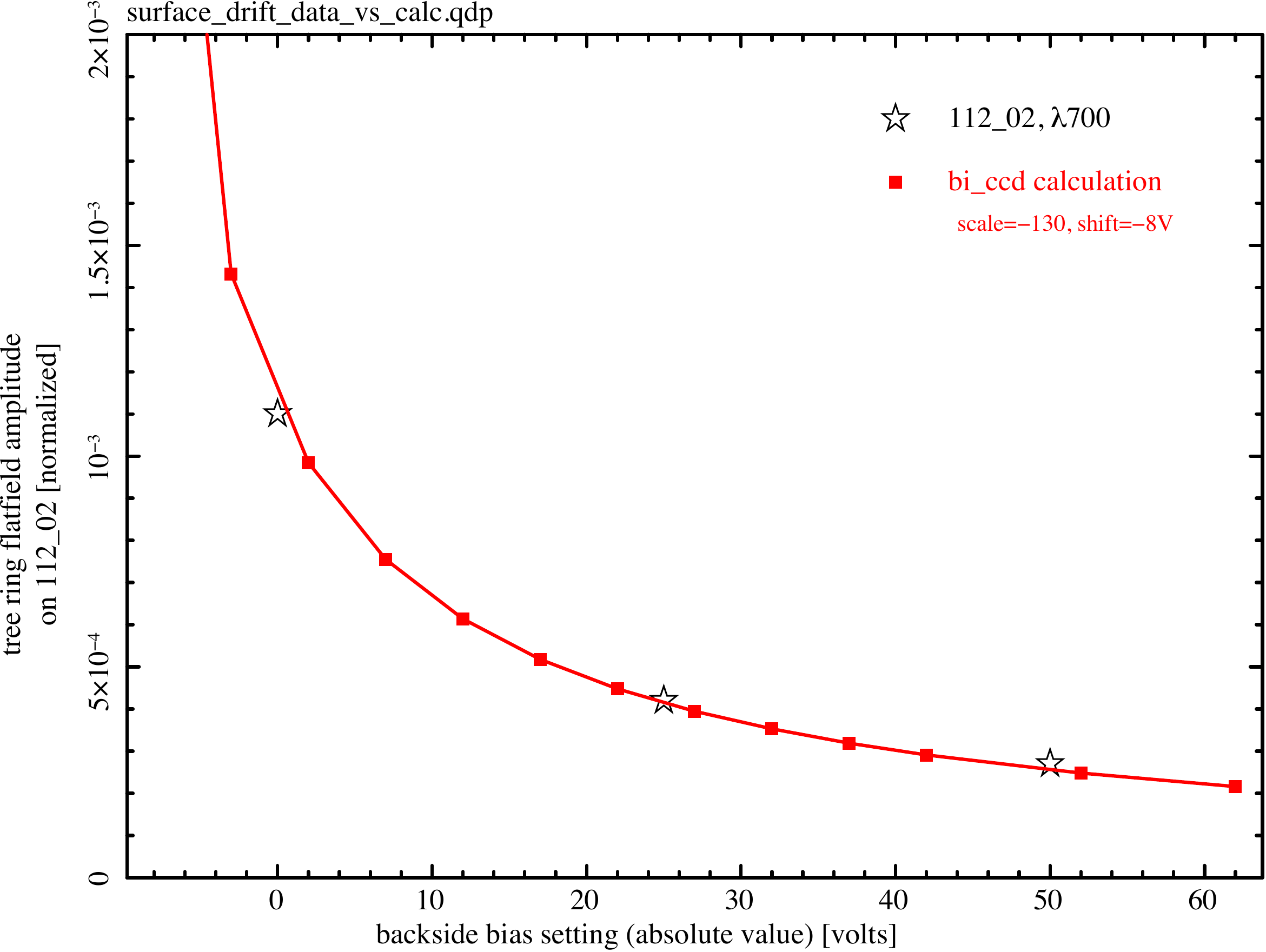}\\
\end{tabular}
\caption{
The fundamental mechanism that we believe give rise to tree ring features in the flat field response. {\bf Left}: Drift calculations for a fixed impurity concentration gradient, for a specific family of {\it backdrop fields} parameterized by the backside bias voltage. The vertical axis reports the lateral drift coefficient in units of 100~mm$^2$/$\Delta \ln(N_a)$ for nominal values of $N_a\sim 10^{12}\,\mathrm{cm}^{-3}$, while the horizontal axis measures the depth ($z$) measured from the front side gate structure. Individual curves are labeled according to the field strength inside the backside surface for 100$\mu$m thick devices. The depth at which the drift lines undergo a maximum angle depend on assumptions of the doping density and on the backside bias. {\bf Right}: an approximate matchup of the surface conversion drift calculation against the observed tree ring feature flat field distortion amplitude. The vertical axis reports the normalized amplitude of the tree ring features, while the horizontal axis is the backside bias setting. Because the tree ring features are so faint, large quantities of data are required to accurately estimate their amplitudes.}
\label{Fig:treering}
\end{figure}

A second test of the {\it backdrop field} model is available, and consists in the observation of a tree ring flat field distortion feature amplitude's dependence on backside bias voltage setting. In contrast to the X-ray diffusion tests described above, the tree ring flat field distortion is a measure of the lateral displacement contribution of the drift calculation. Our straw man model for the tree ring phenomenon is a simple, fossil impurity density variation that is imprinted onto the silicon wafer from the fabrication of the ingot. We only consider these concentrations to vary with lateral coordinate, and not with depth, or distance from the backside entrance window. The expression we use for the lateral field contribution is $\vec{E}_\perp(z)$\cite[\S4.1, \P3]{Rasmussen_paccd_2014}, which is in turn constructed from one-dimensional solutions to Poisson's equation for the potential $\phi(z)$ that also match the boundary conditions. A family of resulting lateral drift coefficient functions that express $\delta\vec{x}_\perp(\vec{x}_0)$ \cite[\S3, Eq3.4]{Rasmussen_paccd_2014}, for a range of backside bias (BSS) settings, are given in Figure~\ref{Fig:treering}. 

The flat field distortion $\Delta I_{i,i+1}$ \cite[\S3, Eq4.7]{Rasmussen_paccd_2014} that results from a weak, sinusoidal variation in impurity concentration with the lateral coordinate is related to the astrometric displacement $\Delta P_{i,i+1}$ \cite[\S3, Eq4.8]{Rasmussen_paccd_2014} by a factor of $k = {2\pi \over \Lambda}$. This relationship is valid as long as $\Lambda$ (the period of the variation) is at least several pixels in length.

Figure~\ref{Fig:treering} also shows the matching job of the predicted flat field distortion $\Delta I_{i,i+1}(BSS)$ to measured tree ring amplitudes, against a handful of data points drawn from a specific location on the device. The predicted distortion was scaled in $y$ (to calibrate the ratio of impurity variation amplitude to its period, $\Delta N_a/\Lambda$) and shifted in $x$ (to calibrate the BSS setting at which the surface field strength is zero) to roughly match up the curve prediction for surface conversions - to the three available data points for 830~nm illumination.  The tree ring flat field distortion amplitudes seen on these devices is certainly small, and just large enough to measure. This validation is somewhat more useful than the diffusion measurement based validation, primarily because the predictions are independent of carrier mobility, while the assumed doping concentration profile is not ruled out by the data.

\section{Fixed pattern features} 
\label{sec:fixedpattern}
Using the {\it backdrop field} validated in \S\ref{subsec:backdrop}, we build up a full drift field $\vec{E}^{tot}(\vec{x})$ by including the image expansion terms that correspond to the channel stop barriers and to the barrier clocks in the perpendicular direction. Mathematical expressions for these terms take the form of $\delta\vec{E}_\perp(\vec{x}|\vec{x}_0)$ \cite[\S4.2, Eq4.10]{Rasmussen_paccd_2014}. 

\subsection{Midline charge redistribution and edge rolloff} 

We test the framework by adding terms one at a time, while comparing to characterization data acquired on these sensors. The cases of the dominant fixed pattern features provide a good sanity check. In the case of the {\it midline charge redistribution}, we position a single, isolated p+ implant feature (essentially identical to individual channel stops) perpendicular to the regular grid of channel stop implants. In the case of the {\it edge rolloff}, we truncate the grid of channel stop implants after the last pixel boundary definition. As a simple example, we do not include a guard ring drain bias outside of the imaging area\footnote{Geometric detail of the guard ring electrode structure is not shared -- for at least one vendor.}.

\begin{figure}
\centering
\begin{tabular}{cc}
\includegraphics[width=0.35\textwidth,angle=0]{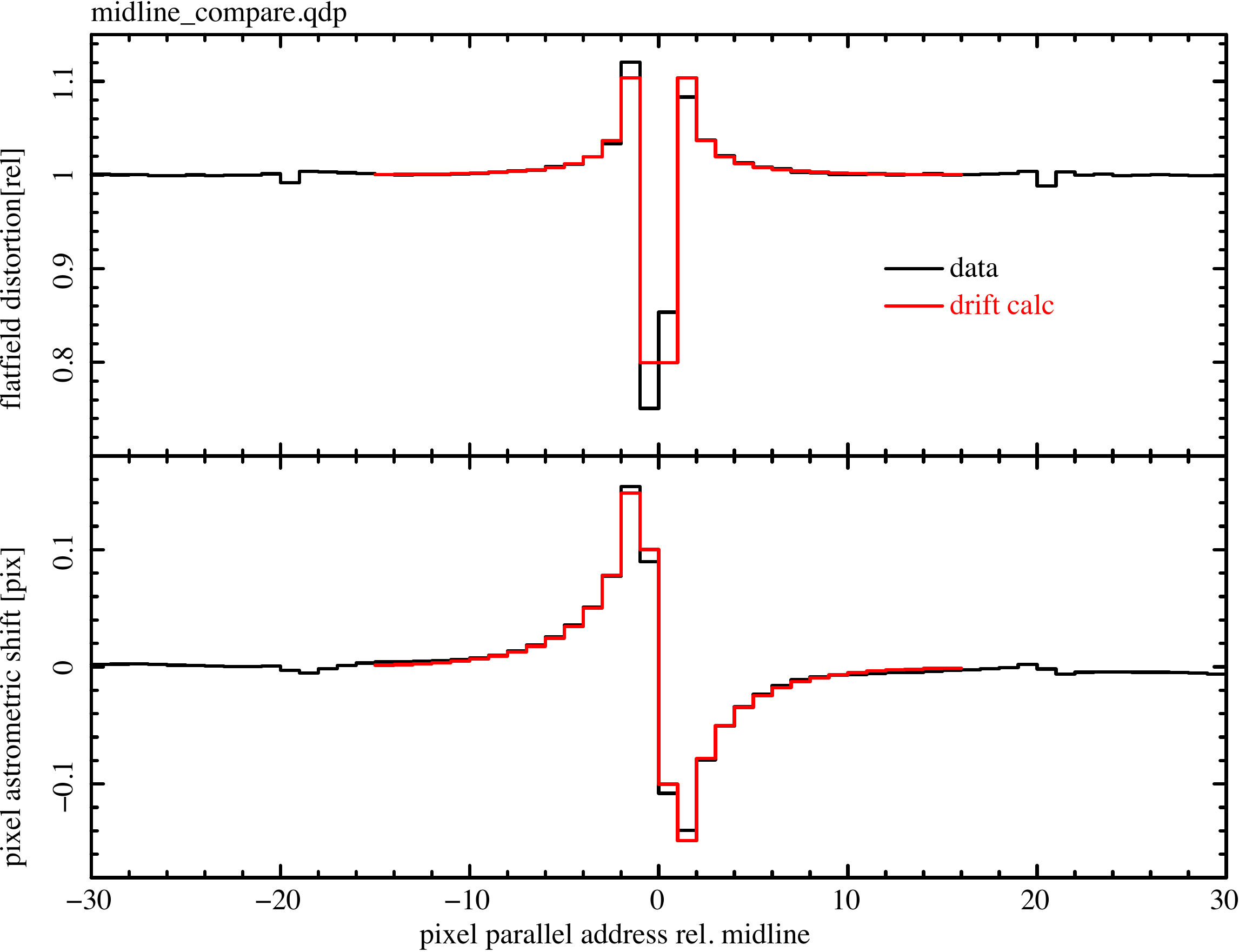}&
\includegraphics[width=0.35\textwidth,angle=0]{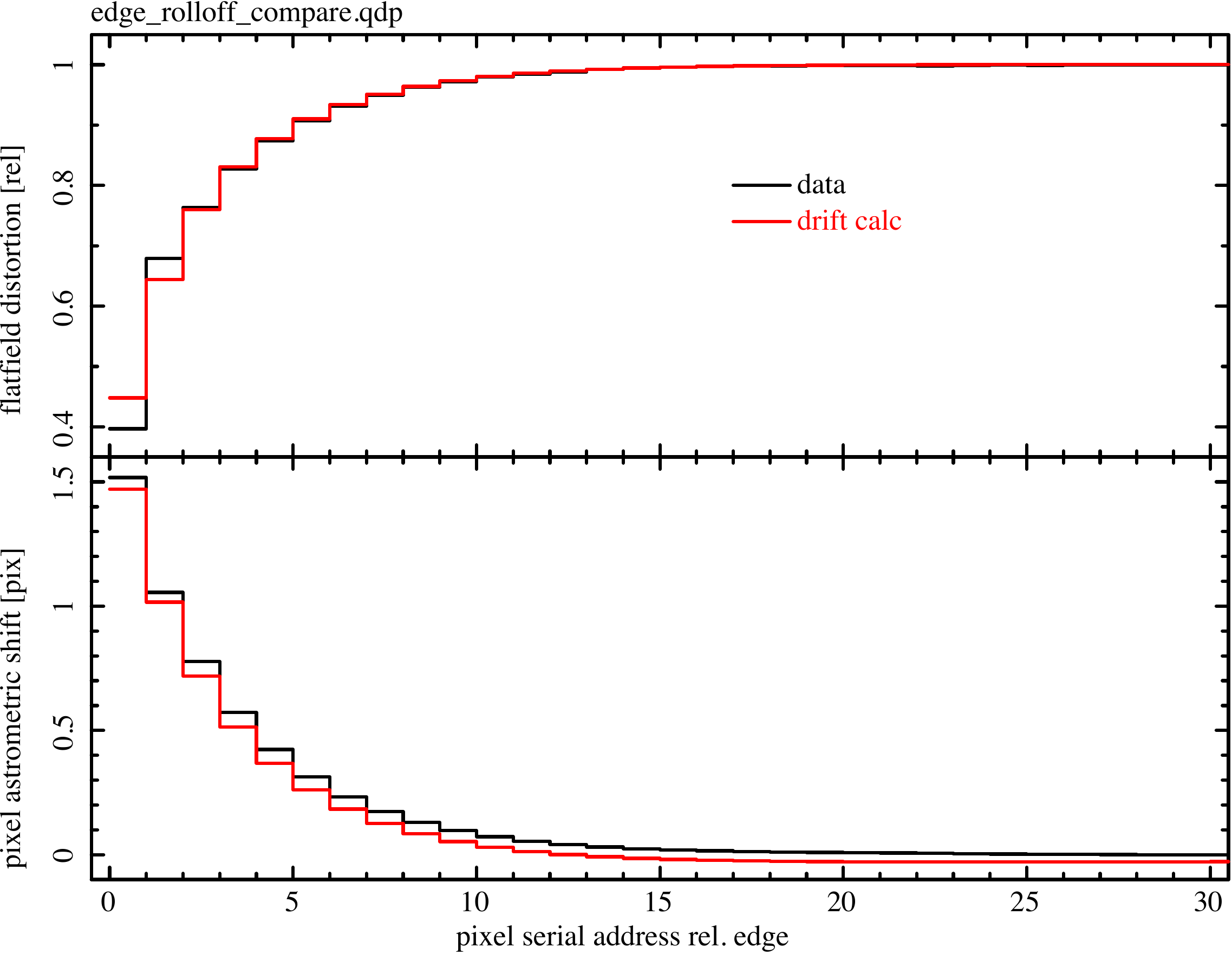}\\
\end{tabular}
\caption{Comparison of flat field response distortions in the data (black curves) to results of the drift calculation (red curves). Each plot displays a comparison of the flat field response (vertical axis) as a function of pixel address (horizontal axis) in the top half - and the pixel astrometric shift (vertical axis) in the bottom half. While the drift calculation has access to the "truth" in reporting astrometric shifts, these shifts are estimated on the data side by assuming an astrometric boundary condition and that pixel response is affected by pixel size only. All dipole moment parameters for the $p+$ implants are quoted in units of $\xi_0$. {\bf Left}: the midline charge redistribution, modeled with $\vec{\xi}_\perp=-7\xi_0\,\hat{k}$. There appears to be a phasing error in the position of the bloom stop with respect to the integrating clocks. {\bf Right}: the edge rolloff, modeled with a truncated array of channel stops, each with $\vec{\xi}_\perp=-28\xi_0\,\hat{k}$ and no guard ring. The difference between the data and model in the first two columns may be most responsive to whether a guard ring drain electrode is modeled. Additional ancillary pixel data resulting from these calculations include pixel shape transfer $\Delta S_{i,i+1}$ and diffusion scale $\sigma(z)$, which is in general position dependent. In the case of the edge rolloff, we find a local maximum in $\sigma(t_{Si})$ at the edge that is about 8\% larger than nominal in the edge-most pixels, corresponding to drift times that are $\sim$17\% greater than nominal.}
\label{Fig:fixedpattern}
\end{figure}

Figure~\ref{Fig:fixedpattern} provides the comparison between the characterization data and modestly tuned instances of the drift calculation, set up in each of the cases as described above. The strength of the two-dimensional dipole moments used ($\vec\xi_\perp$) are noted in the caption. These comparisons are provided only for a single backside bias setting in the data. Although we have not yet tested the fidelity of this drift calculation for the range of backside bias voltages available in the data, we have reasonable confidence that these will compare nicely - primarily because of our {\it backdrop field} validation efforts of \S\ref{subsec:backdrop}.

\subsection{Bamboo} 
\begin{figure}
\centering
\begin{tabular}{cc}
\includegraphics[height=0.27\textwidth,angle=0]{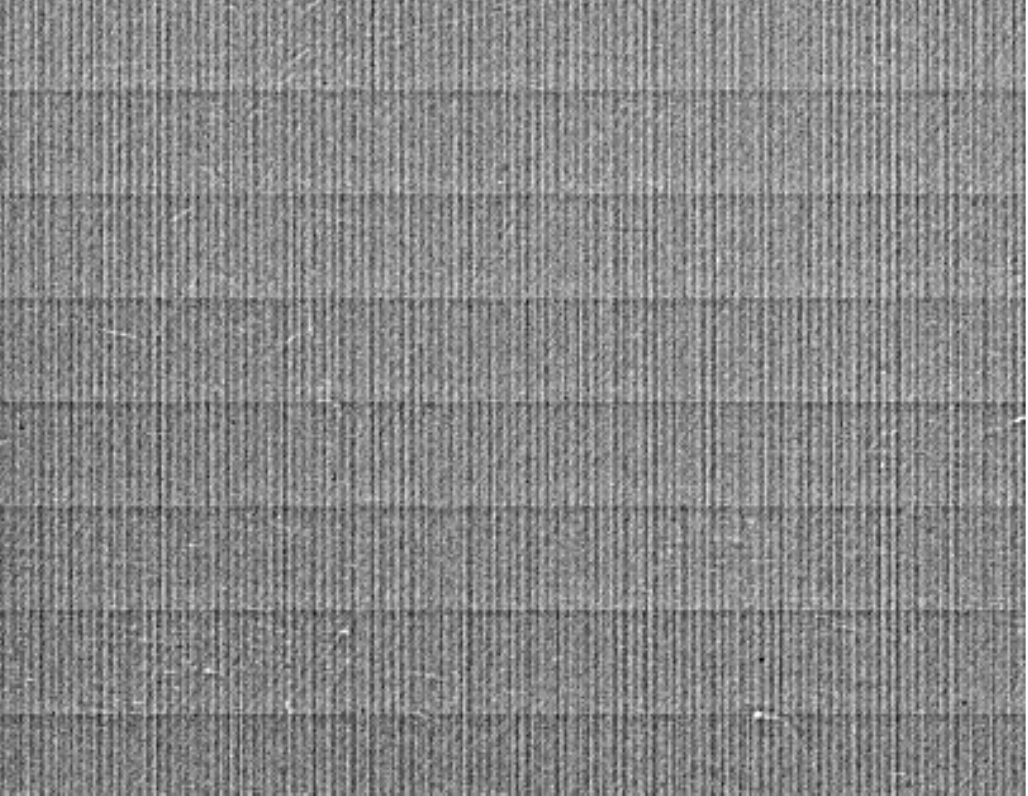}&
\includegraphics[height=0.27\textwidth,angle=0]{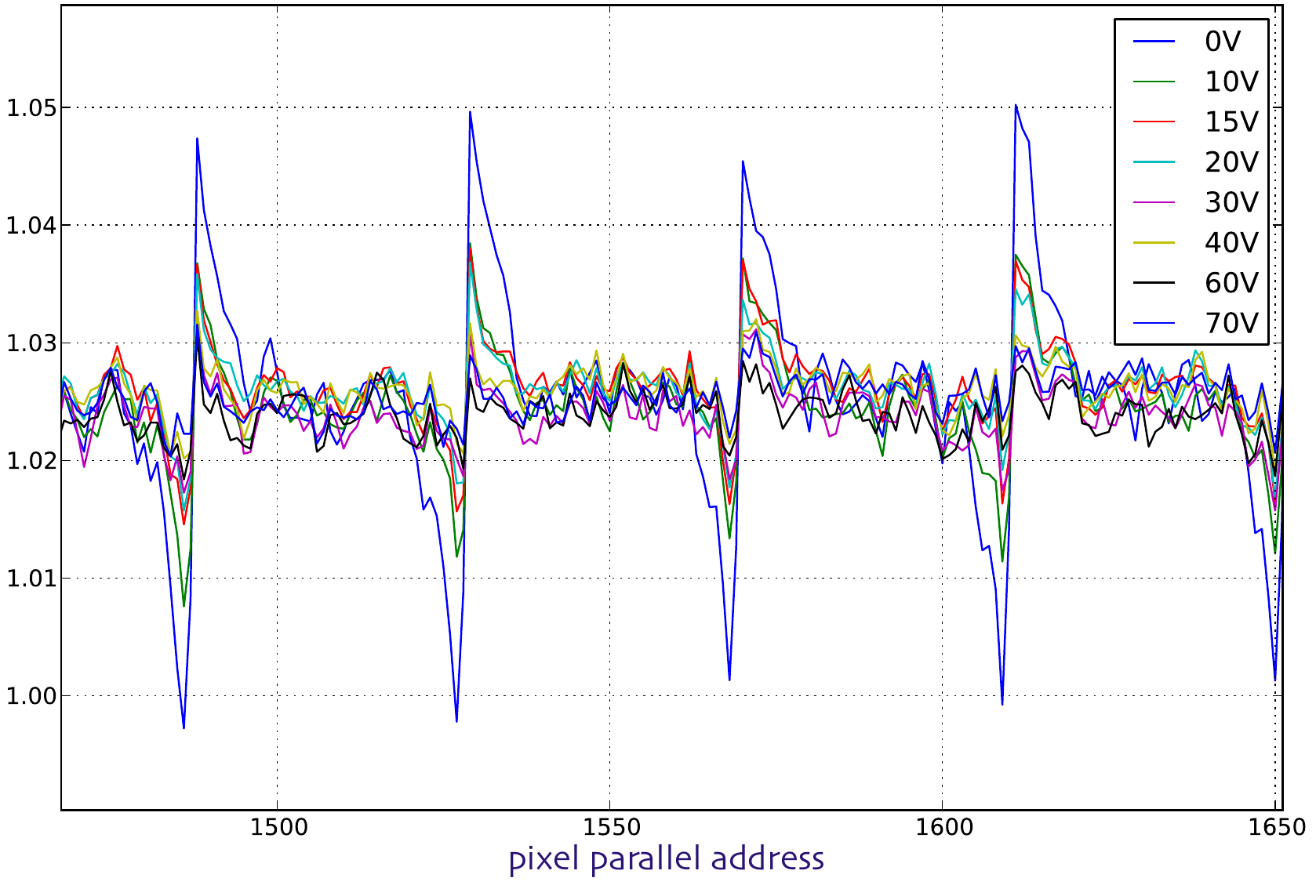}\\
\multicolumn{2}{c}{\includegraphics[width=0.7\textwidth,angle=0]{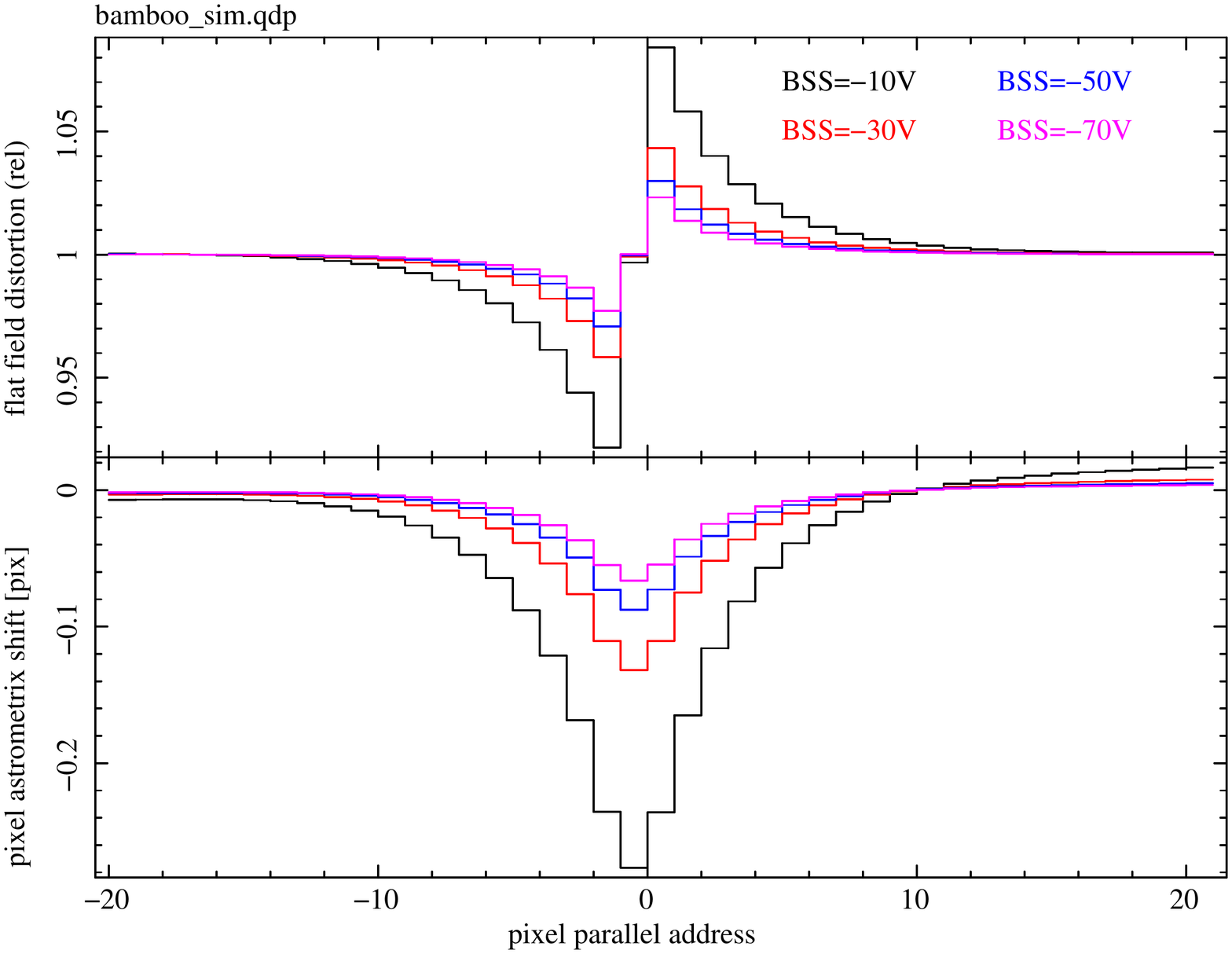}}\\
\end{tabular}
\caption{
{\bf Upper left}: The {\it bamboo} flat field distortion feature is seen with a 41 pixel (410$\mu$m) period along the (vertical) parallel pixel address axis, with a less coherent, higher frequency distribution along the (horizontal) serial axis. The normalized amplitude of this feature is approximately $\Delta \ln(C) \sim 0.05$.  {\bf Upper right}: the periodic fluctuation in flat field response (vertical axis) is seen as a function of parallel address (horizontal axis). The distortion reduces as the backside bias voltage is adjusted, to increase the electric field strength inside the backside window. The amplitude apparently reduces by a factor of 5 or so as $|$BSS$|$ is raised from 0 to 70V. {\bf Bottom}: the drift calculation is performed and the {\it bamboo} is modeled as small, but abrupt change in the channel stop implant concentration as a function of the pixel parallel address coordinate. The top plot displays the flat field response (vertical axis) as a function of parallel address (horizontal axis), together with the pixel astrometric shift (vertical axis, bottom plot). The magnitude of the step function used here is relatively small: $\Delta \ln(|\vec\xi|) \sim 0.03$ as one crosses the ``zero'' mark in the pixel coordinate system shown. The strength of the flat field distortion feature in the calculation also drops by a factor of 3 or so as $|$BSS$|$ is raised from 10 to 70V.
}
\label{Fig:bamboo}
\end{figure}

As mentioned above, the flat field distortion feature we refer to as {\it bamboo} fortunately is not seen in all sensors, and its presence has been traced to a new e$^-$ beam lithographic process that will be discontinued. Regardless, its properties can still be understood in the context of the drift calculation. The resemblance between the calculation and data is surprising. The periodic, bimodal distortions seen should apparently be accompanied by a localized ($\sim$7 pixel FWHM), correlated pixel astrometric error with peak errors on the order of 0.25~pixels. The fact that the flat field distortion is bimodal does not favor a simple geometric step-and-repeat error, but probably a $\sim$3\% non uniformity p+ implant concentration (or equivalently, a nonuniform implant depth) that is repeated over the 410$\mu$m period. See Figure~\ref{Fig:bamboo} for more details. We do not have an adequate explanation or model that would explain the high frequency variation along the pixel serial address axis, which appears to be incoherent as compared to the distortions projected along the parallel address axis.

\subsection{Anti-correlation of adjacent pixels} 

We believe the drift calculation may be used to better understand the underlying mechanism of the pixel-to-pixel anti-correlations studied in deep PRNU images\cite{Smith:2008,Kotov:2010} particularly along the serial address axis (column-to-column). We have not yet devised a prediction, expressed in the sort of observables used to study this phenomenon, that would discriminate between simple geometric pixel boundary jitter and channel stop dipole moment strength variation induced partition noise. The latter would predict pixel boundary distortions that affect pairs of pixels at a time (with the same sign), rather than isolated, adjacent pixel pairs that show flat field response anti-correlation. Previous studies of the {\it tearing onset}\cite[\S4.4, Figs5 \& 3d]{Rasmussen_paccd_2014} show the sort of the {\it Greens function} unit contribution to the PRNU we would consider. This would be an interesting topic to pursue at some point. 

\section{Dynamic terms} 
\label{sec:dynamicterms}
The drift calculation naturally can provide pixel boundary distortion predictions due to collected conversions at the channel\cite[\S4.5]{Rasmussen_paccd_2014}. At least two types of observables are available to constrain the model parameters of the drift calculation: These include direct measurements (e.g., the {\it brighter-fatter effect}) and the inferential methods (the {\it mean-variance relation} and {\it flat-field autocorrelations}). A general problem with reliance on direct measurements {\it alone} lies with limited understanding and control of absolute properties of a focused spot, and such investigations can rapidly become circular if they are not conducted in a series of well controlled exposures made in rapid sequence\footnote{Fine dithering control and/or illumination with a calibrated pinhole array can potentially circumvent this degeneracy.}. Inferential methods, in contrast, rely on the fundamental stochastic assumption that local incident flux follows a Poissonian probability distribution. Assumptions that pixel areas are identical or that pixel areas are constant with flux - are specifically not made. Direct and inferential methods used together provide enhanced diagnostic power, and we expect to identify efficient characterization methods for ancillary pixel data retrieval as an outcome.

\begin{figure}
\centering
\includegraphics[width=0.5\textwidth,angle=0]{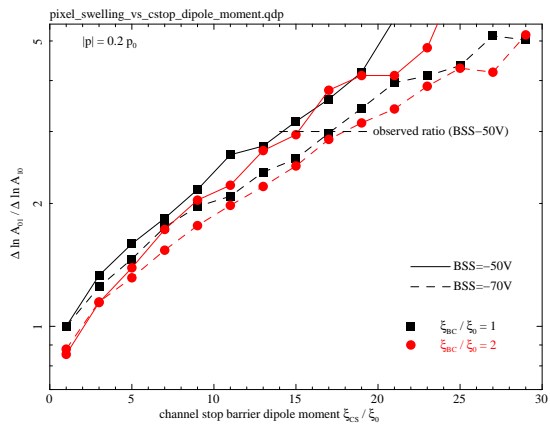}\\
\caption{
Dependence of the adjacent pixel area deformation response to channel content, as a function of channel stop barrier dipole moment. The vertical axis reports the response ratio ($\Delta \ln A_{01}/\Delta \ln A_{10}$) - which is observed to be close to 3 - as a function of channel stop barrier dipole moment ($\xi_{CS}/\xi_0$) as the horizontal axis. The four curves shown are given to show adjacent behavior for a doubling of the barrier clock dipole moment ($\xi_{BC}/\xi_0$) and for an adjacent backside bias setting. Interestingly, the ratios computed here cannot be reproduced by considering only the deformed pixel width along the plane that includes the aggressor (conversions within in the channel). The ratios are as large as they are due to the fact that the pixel ``height'' in the transverse direction shrinks, even as the ``width'' increases while its centroid shifts toward the occupied channel.
}
\label{Fig:chanstoptune}
\end{figure}

Starting with the observation\cite[Fig.2]{Antilogus_paccd_2014} that autocorrelation terms $A_{01}$ and $A_{10}$ apparently differ by a factor of $\sim$3 (under a certain operating condition), we search for channel stop and barrier clock dipole strengths that would cause pixels with ($\Delta x,\Delta y$)=(0,1) to increase in area $3\times$ faster than for the pixel with ($\Delta x,\Delta y$)=(1,0) - in response to a finite collection of conversions contained within the pixel's channel. Figure~\ref{Fig:chanstoptune} provides a demonstration that autocorrelation results can readily be used to constrain and validate the drift calculation. This of course implies that the drift calculation can be used to generate a validated Greens function that can in turn be used to forward fold and predict the detailed distortions of pixel boundaries as images are integrated.

\begin{figure}
\centering
\begin{tabular}{ccc}
\includegraphics[width=0.25\textwidth,height=0.25\textwidth,angle=0]{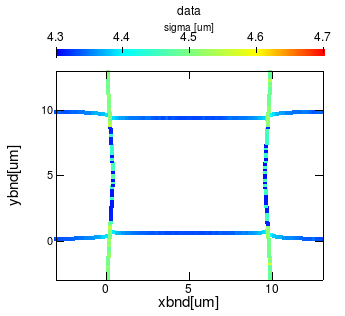} &
\includegraphics[width=0.25\textwidth,height=0.25\textwidth,angle=0]{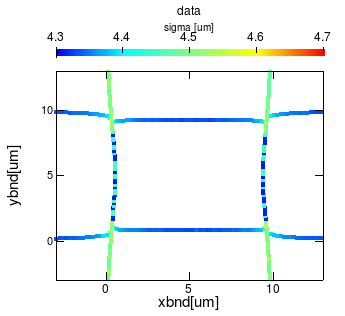} &
\includegraphics[width=0.25\textwidth,height=0.25\textwidth,angle=0]{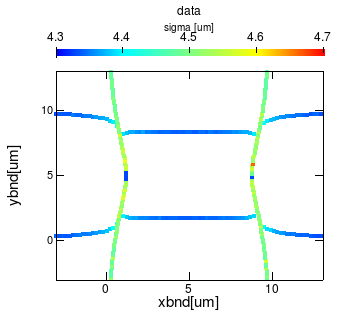} \\
\includegraphics[width=0.25\textwidth,height=0.25\textwidth,angle=0]{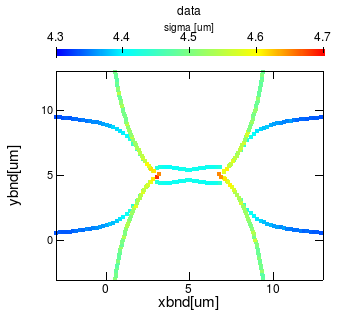} &
\includegraphics[width=0.25\textwidth,height=0.25\textwidth,angle=0]{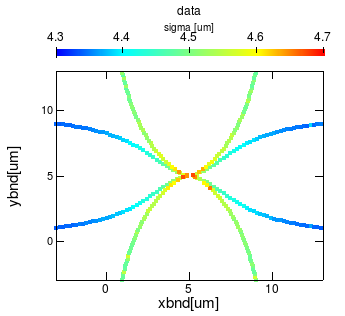} &
\includegraphics[width=0.25\textwidth,height=0.25\textwidth,angle=0]{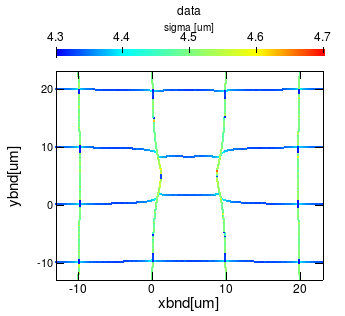} \\
\end{tabular}

\caption{
Pixel boundary distortion Greens functions that correspond to barrier clock and channel stop dipole moments determined using the method shown in Figure~\ref{Fig:chanstoptune}: $(\vec{\xi}_{\perp CS},\vec{\xi}_{\perp BC}) = (-15\xi_0\hat{k},-1\xi_0\hat{k})$ for a backdrop field corresponding to $|$BSS$|$=50V. 
For each plot, the pixel boundaries are reported in microns along the parallel (vertical) and serial (horizontal) axes.
Large values were used for the isolated channel content dipole moment (seen here centered at $x=+5\mu$m, $y=+5\mu$m) so that the distortions are clearly visible. From the {\bf top}, {\bf left} to {\bf right}, the channel content dipole moment ($\vec{p}$) settings were $-1p_0\hat{k}$, $-2p_0\hat{k}$, $-4p_0\hat{k}$, $-8p_0\hat{k}$ and $-16p_0\hat{k}$. The plot on the {\bf lower right} is a zoom-out of the plot directly above it ($\vec{p}=-4p_0\hat{k}$). In each plot, the color of the plotting symbols encode the lateral diffusion computed for conversions that occur directly over the pixel boundaries.
}
\label{Fig:pixelboundaries}
\end{figure}

Figure~\ref{Fig:pixelboundaries} provides detailed examples of the Greens function we compute using the tuned model for the pixel barrier dipole moments and for a standard {\it backdrop field}. To generate these we have exaggerated the strength of dipole moment that represents collected conversions. We have not yet simulated a full flat field response autocorrelation function while using these full Greens functions; we have only performed this for simple boundary translations in response to pixel content (which does in fact nicely reproduce leading terms of the autocorrelation matrix). This simulation would be a necessary validation step because the autocorrelation matrix can potentially contain tremendous detail, with a high signal-to-noise.

\section{Conclusions} 
We have put forward a modeling framework that will assist us in our estimation of instrumental imaging systematics that will be present in on-sky data for future dark energy experiments. This framework is based on first principles and electrostatics, and appears to show some promise when it comes to our quantitative interpretation of sensor characterization data. Quantitative matches to fixed pattern features in flat field response would suggest that significant fractions ($\sim$5\%) of the focal plane pixels may be retained and analyzed rather than trimmed off and discarded - and this would require use of {\it ancillary pixel data} that encode astrometric shifts, pixel area, pixel shape transfer, and perhaps even position dependent isotropic diffusion\cite{Rasmussen_paccd_2014}. Although it was not discussed or shown in this paper, the depth dependence of the pixel boundary is also available\footnote{This is a natural byproduct of the drift calculation.} and this can be used to compute wavelength- or SED-dependence of these pixel systematics. 

Results of these calculations will be used to encode these systematics into an efficient pixel partition library code that can in turn be invoked from either LSST's photon simulator\cite{Peterson_2014} or any other image simulator code to study the impact of pixel response distortion effects, one mechanism at a time. At present, we do not yet know the impact of these unmitigated systematic effects on the science yield for LSST in the context of a full survey with hundreds of viewing instances for each field. 
We are engaging LSST's Dark Energy Science Collaboration (DESC) for assistance in answering these questions, and if necessary, will (together with LSST's data management team) identify algorithms that can utilize the ancillary pixel data to reduce their impact.

Because the DECam is already in operation and {\it is} affected by the sorts of systematics we model, we plan to adapt parameters of this framework to perform drift calculations that may be validated against similar, measurable quantities of the DECam sensors, and a testbed for astronomical analysis using such ancillary pixel data may be exercised.

\label{sec:conclusions}


\acknowledgments     
We gratefully acknowledge Tony Tyson for providing an LSST internal review of this paper prior to its submission to SPIE. The comments and suggestions he provided led to a substantially improved manuscript.

LSST project activities are supported in part by a Cooperative Agreement with the National Science Foundation managed by the Association of Universities for Research in Astronomy (AURA), and the Department of Energy. Additional LSST funding comes from private donations, grants to universities, and in kind support from LSSTC Institutional Members.

\bibliography{spie-astro-tel-instrum_2014_arasmus}   
\bibliographystyle{spiebib}   

\end{document}